\documentstyle[prl,aps,multicol,epsfig]{revtex}
\renewcommand{\narrowtext}{\begin{multicols}{2} \global\columnwidth20.5pc}
\renewcommand{\widetext}{\end{multicols} \global\columnwidth42.5pc}
\def\top#1{\vskip #1\begin{picture}(290,80)(80,500)\thinlines \put(
65,500){\line( 1, 0){255}}\put(320,500){\line( 0, 1){
5}}\end{picture}}
\def\bottom#1{\vskip #1\begin{picture}(290,80)(80,500)\thinlines \put(
330,500){\line( 1, 0){255}}\put(330,500){\line( 0, -1){
5}}\end{picture}}

\begin{document}
\newcommand{\be}{\begin{equation}}
\newcommand{\ee}{\end{equation}}  
\newcommand{\ba}{\begin{eqnarray}}
\newcommand{\ea}{\end{eqnarray}}
\title{Ground State Wavefunctions of General Filling Factors in the Lowest
          Landau Level}
\author{J. H. Han$^{1}$ and S.-R. Eric Yang$^{1,2}$}
\address{Asia Pacific Center for Theoretical Physics, 207-43
              Cheongryangri-dong Dongdaemun-gu Seoul 130-012, Korea$^{1}$
                  \\ and \\
             Department of Physics, Korea University, Seoul 136-701,
               Korea$^{2}$}
\maketitle
\draft
\begin{abstract}
We present a set of explicit trial wavefunctions for the filling factors
$\nu=n/(2n\pm
1)$ and $\nu=1/2$ in the symmetric gauge. We show that the zeroes of the
wavefunction, except those
dictated by the Fermi statistics, are detached from the particles. The evolution
of zeroes as the filling factor is varied is examined. We show that the
wavefunction at half-filling exhibits a $2k_F$-like oscillation in its 
occupation number profile. The center-of-mass motion of the ground state
droplet is described in terms of the intra-Landau-level excitations of
composite fermions. 
\end{abstract}
\narrowtext

The remarkable discovery of the fractional quantum Hall effect has led to
a rich variety of beautiful new concepts in the condensed matter
physics of low-dimensional systems\cite{books}. The essential
ingredient in the description of the fractional Hall systems is the
Jastrow correlation, with which one can write down the
eigenstates in the lowest Landau level (LLL) as\cite{drop}
\be
\psi_{LLL}=
P_{L^3} [\prod_{i<j}(z_i-z_j)^{2} \phi_{\nu^*}].
\label{eq:CF_waveftn}
\ee
This wavefunction describes states whose filling factors are centered around
$\nu=1/2$. One obtains $\nu=1/3, 1/2$, and in general
$n/(2n\pm 1)$, by choosing a filled Landau level (LL), a Fermi gas, and 
$n$-filled LLs  for $\phi_{\nu^*}$
respectively\cite{laughlin,jain,rr} and projecting to the LLL ($P_{L^3}$).
Equation (\ref{eq:CF_waveftn})
does not, however, give us the analytical form of $\psi_{LLL}$ after
the projection.


Recently, Jain and Kamilla (JK) proposed wavefunctions lying entirely in the 
LLL without the projection\cite{JK,HG}. They are based on the 
``composite fermion (CF) wavefunction", obtained from the
corresponding electronic state $\psi^{e}(z_i)$ by 

\be
    \psi^{CF}(z_i)=P_{L^3}[\psi^{e}(z_i,\bar{z}_i)J_i], 
\label{eq:CF_state}
\ee 
where
$J_i$ equals $\prod_{k\neq i}(z_i-z_k)$ in the symmetric gauge.  
$\psi^{CF}(z_i)$ is not a genuine single particle state
as its value depends on the position of all the other particles.
Unlike in Eq.\ (\ref{eq:CF_waveftn}), the long-range Jastrow correlation 
is built in at
the level of individual 
wavefunctions. The ground states of interacting
fermions in the LLL for a given total angular momentum $M_{tot}$ are of the 
general determinant form
   
\be
\left|\begin{array}{cccccccc}
J_1 & z_{1} J_1 & ... & z_{1}^{N_0 -1}J_{1} & \partial_1 J_1 &... &
z_{1}^{N_1 -1}\partial_1 J_{1} & ... \\
J_2 & z_{2} J_2 & ... & z_{2}^{N_0 -1}J_{2} & \partial_2 J_2 &... &
z_{2}^{N_1 -1}\partial_2 J_{2} & ... \\
  & & & ... & & & &     \\
J_N & z_N J_N & ... & z_{N}^{N_0 -1}J_{N} & \partial_N J_N & ... &
z_{N}^{N_1 -1}\partial_N J_{N} & ...
     \end{array}\right|.
\label{eq:generalwaveftn}
\ee
Successively higher powers of $\partial_i$ acting on $J_i$ appear as we
move to the right for a given row. The matrix elements are of the
form $z_i^{m}\partial_{i}^{n}J_i$. One may regard $n$ and $m$ as
the LL index and the angular momentum of the CFs respectively. $N_i$ is
the number of CFs occupying the
$i$-th LL. For given $N_i$, only the compact configuration,
$m=0,1,..,N_i -1$, occurs in the ground state.
We will adopt JK's notation $[N_0, N_1,..,N_{l-1}]$ to denote a given
state.
The sum $\sum_{i=0}^{l-1}N_i$ adds up to the total particle number $N$.
In this notation, the $1/3$ Laughlin state is given by $[N]$; only the LLL
state is filled. The filled electronic Landau level $\nu=1$ is given by
$[1,1,..,1]$. No specific prescription was given for other general filling factors.

In physical terms, the Jastrow factor $J_i$ implies that each particle
carries
with it a correlation hole with respect to all the other particles. The
projection $P_{L^3}$ is tantamount to replacing the anti-holomorphic
variable $\bar{z}$ with $2(\partial/\partial z)$\cite{gv}. Some of the
correlation zeroes, both in the compressible and incompressible states,
must be displaced since $\partial/\partial z$ acting on  
the Jastrow factor will in general displace its zeroes.
It is believed that the resulting particle-hole pair makes a dipole-like
entity. The field theory of these dipole structures is a subject of active
study in recent years\cite{read,recent}. The Laughlin  
state contains no anti-holomorphic part, hence all the zeroes are tightly
bound on the particles\cite{acta}.
For other fractions, the interplay between
zeroes and the particle coordinates in the system is not as clearly
understood. 

We propose a class of ground state
wavefunctions with the filling factor $\nu=n/(2n\pm 1)$, including
$\nu=1/2$, in the symmetric gauge using Eqs.\
(\ref{eq:CF_state}) and (\ref{eq:generalwaveftn}). 
They are valid for a symmetric confining potential which preserves the
angular momentum $M_{tot}$.
The filling factor is defined by  
$\nu=N(N-1)/2M_{tot}$ \cite{girvin,trugman}, where $N$ can be small
(quantum dot) or a macroscopically large number.
Conceptually our
wavefunctions are based on the flux attachment scheme
that binds two flux quanta to each electron, but differ from the conventional CF
picture  in the way the residual flux are treated.
For example, at $\nu=2/5$, one is left with 1/2 flux quanta per particle.
The standard CF
picture is obtained by assuming this residual flux to form a $\nu^* =2$
state. However, one can also regard the residual flux as three  
integer flux quanta plus an extra one pointing in the {\it opposite
direction}, shared between four particles. As a result, three-quarters of
particles that bind to upward flux lie in the  holomorphic LL while the other
one-quarter lies in
the anti-holomorphic LL\cite{comment}. The anti-holomorphic wavefunctions become
derivatives acting on $J_i$ with the usual substitution
$\bar{z}_i^m
\rightarrow \partial_i^m$. According to this picture  $\nu=2/5$ is
given by $[3N/4,1,..,1]$, which is the case as shown below. The same interpretation
applied to other fractions indicate that the ground state is given by 
$[(\nu^{-1}-1)N/2,1,..,1]$.
At $\nu=1/2$, once the initial flux
attachment is made, half the particles bind to an additional flux in the  
``up" direction while the other half bind to ``down" flux quanta, leading
to a symmetric population of $z_i^m$ and $\partial_i^m\,(\equiv
\bar{z}_i^m)$ in the wavefunction.

Following these ideas, we explicit construct the ground state wavefunctions at
$\nu=2/5, 3/7, 3/5, 1/2$, and $2/3$ and list the results in Table I. The
states at half-filling, $M_{tot}=N(N-1)$, were discussed
previously\cite{yang}. 
The total angular momentum of $\nu=2/5$ state, for example, is  
$M_{tot}=5N(N-1)/4$. In order to have an integer-valued 
$M_{tot}$, $N$ is either 
a multiple of 4, $4k$, or $4k+1$. One can check that the desired 
angular momentum is obtained for

\ba
      N=4k+1: & [3k+1,1,..,1], \nonumber \\
      N=4k:   & [3k,2,1,..,1].
\ea
For $N\le 7$, our constructions give the correct ground states
at the given angular momentum $M_{tot}$\cite{JK}.

We can make general observations about the wavefunction.
First, the states are given by $N_0 /N=(\nu^{-1}-1)/2$ for large $N$,
$N_1 =1$ or $2$ depending on the parity of the particle number, and 
$N_i=1\, (i\ge 2)$, largely in agreement with the argument presented
above. 
There exists an interesting dual 
relation between states $\nu=n/(2n+1)$ and $\nu=n/(2n-1)$ for given $n$.
Starting from one state, we can obtain the other by interchanging
$z_i^{m}$ with $\partial_i^{m}$ in the determinant, Eq.\
(\ref{eq:generalwaveftn}). 
This is not the same as the particle-hole symmetry which relates two
states the
sum of whose filling factors is one\cite{girvin}. 
One has  $1/n$ residual flux per particle  pointing either upward or down 
at $\nu=n/(2n\pm 1)$. Our argument for the assignment of the residual flux
goes through in exactly the same manner for both states, with the role of the
holomophic and the anti-holomorphic parts reversed. In the projected space, this
leads to the duality under the interchange of $z_i^{m}$ with $\partial_i^{m}$.

\begin{figure}[ht]
\centering
\leavevmode
\epsfig{file=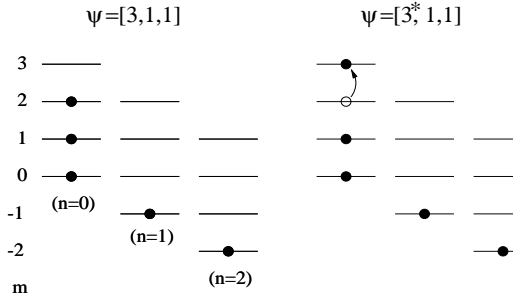,height=4cm,width=7cm}
\caption{A schematic picture of the ground state
$[3,1,1]$ and the excited state $[3^*,1,1]$ in terms of CF
levels.}
\label{fig:schematic}
\end{figure}

The low-energy excitation spectrum of the states is easily classified 
within the CF picture. One can distinguish between intra-Landau-level
excitations which preserve $n$, and the inter-Landau-level excitations
which change the LL index. An example of the intra-Landau-level,
$m\rightarrow m+1$ excitation is shown in Fig.\ \ref{fig:schematic}.
For $\delta M_{tot}=1$, we have $l$ distinct
intra-Landau-level excitations if the ground state consists of $l$
different LLs, $[N_0,..,N_{l-1}]$. Such a plethora of low-energy states is 
difficult to understand for incompressible fractions, where the low-energy 
modes are known to be one-dimensional edge excitations\cite{cll}. We shall 
return to the resolution of this puzzle later after we have explained the 
presence of the ``occupation tail" in our wavefunctions.

The center-of-mass (CM) motion of a droplet is described by multiplying the
droplet wavefunction with powers of the CM coordinate
$Z=(\sum_{i=1}^{N} z_i)/N$\cite{trugman,macdonald}. 
In the absence of the external  
potential, the
displaced state is degenerate in energy with the original state because 
the interaction Hamiltonian, projected to the LLL,
commutes with $Z$\cite{trugman}:
\be
\sum_{k,i<j} V(k)[e^{i\bar{k}(z_i-z_j)/2}
 e^{ik(\partial_i-\partial_j)},Z]=0.
\ee
We have found that $\psi^{\ast}=Z\psi_{g.s.}$ has a remarkably simple
interpretation in terms of the CF picture. Since
$\psi^\ast$ is degenerate with $\psi_{g.s.}$ itself, it is natural to
expect
$\psi^\ast$ to be expressed as a linear combination of the
intra-Landau-level excitations of CFs. We denote
$[N_0,..,N_{i}^{\ast},..]$ as the state where the highest angular momentum
of the $n=i$ LL has increased by unity (See Fig.\
\ref{fig:schematic}). We have
explicitly confirmed for $\nu=2/5\,\, (N=4)$, $\nu=1/2\,\, (N=3,4,5)$ and
$\nu=2/3\,\, (N=4,5)$ that the following relation holds;

\be
\psi^{\ast}=[N_0^{\ast},..]+[N_0,N_1^{\ast},..]+\cdots+
            [..,N_{l-1}^{\ast}].
\label{eq:CMmode}
\ee
Note that each bracketed state is not
normalized but rather is a Slater determinant of Eq.\ 
(\ref{eq:generalwaveftn}) with the overall constant
equal to one. Equation (\ref{eq:CMmode}) obviously holds in the special
limit of $\psi_{g.s.}=[N]$\cite{cll}.

\begin{figure}[ht]
\centering
\epsfig{file=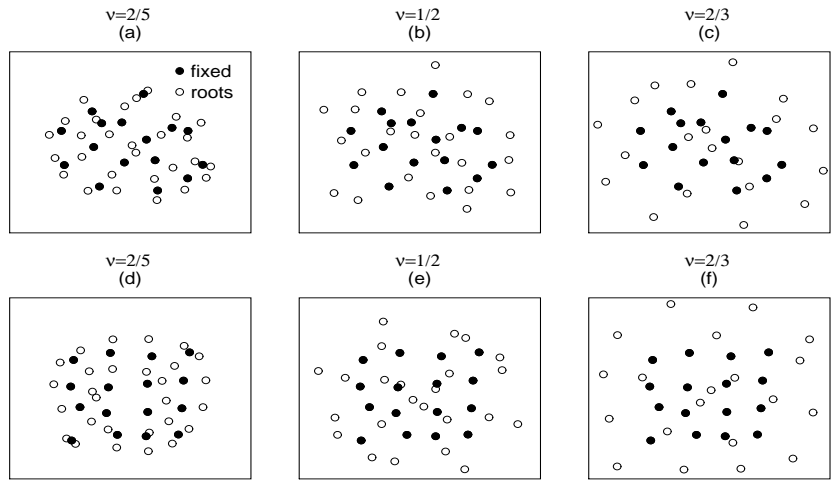,height=12cm,width=8cm,angle=270}
 \caption{A plot of zeroes ($\circ$) for a random ((a)-(c)) and a
quasi-regular ((d)-(f)) distribution of particles($\bullet$) for
filling factors $\nu=2/5,1/2$, and $2/3$, and $N=17$.}
\label{fig:zeroes}
\end{figure}

We can gain a lot of insight about the nature of the ground states by
plotting the zeroes of the wavefunction, $\psi(z_1, z_2,..,z_N)=0$, with
respect to $z_1$ while the other coordinates $z_2$ through $z_N$
are held fixed. At least one zero occurs at
each fixed position itself due to the exclusion principle, hence our
task lies in finding out where the extra zeroes (those not dictated by Fermi
statistics)  are. 
We have derivatives of $J_i$ entering the determinant, whose zeroes do not
in general occur at $z_i=z_j$ but somewhat displaced from it. We
expect
that the full many-body state will also share this feature. 
Figure \ref{fig:zeroes}(a)-(f) shows a plot of zeroes for two different 
distribution of $z_i$, $i=2,..,N\, (N\!=\!17)$, for several filling factors.
The total
number of zeroes that occur is the same as the degree of
$\psi(z_1,..,z_N)$ treated as a polynomial in $z_1$. We denote this
quantity as $m_{eff}$ and show its value in Table I. One sees that
$m_{eff}/N=(3+ \nu^{-1})/2\, (\nu<1)$ for large systems.

Some qualitative observations can be made about the evolution of zeroes
shown in Fig.\ \ref{fig:zeroes}. As we move from $\nu=1/3$ to $2/5$, two
out of three
zeroes detach themselves and form a (two vortices $+$ one particle)
composite. Since there are fewer zeroes per particle, some
of the composites have to share their zeroes. As we
increase the
density, some of the zeroes  move away from the particle cluster
while the density of zeroes in the interior goes down. At
$\nu=2/3$, we have a very distinct ``halo" of zeroes surrounding the
particles. At $\nu=1$ both the halo and the
interior zeroes disappear completely. We have also observed that the
number of extra zeroes which appear inside the particle cluster is roughly
given by $N_0$, the number of CFs in the LLL. For example, one-quarter of
the total particles reside in
the LLL for $\nu=2/3$, and indeed we have $16\times
(1/4)=4$ zeroes in the cluster's interior in the plots shown in Fig.\
\ref{fig:zeroes}. The exception is the $\nu=1/3$ state, where the outer
zeroes merge with the particles. 
A naive definition of the filling factor as the number of zeroes per
particle will give a lower $\nu$ than what we obtain for $\nu$
in terms of $M_{tot}$. 
In other words, our trial wavefunction carries more zeroes than 
expected in the  thermodynamic limit without the boundary.
\widetext
\top{-3cm}
\bottom{-2cm}
\narrowtext

The over-abundance of zeroes and their tendency to be located outside the
cluster can be related to another feature of the trial wavefunction,
the long tail in the occupation number\cite{yang}. One can 
verify that
$m_{eff}$ is the maximum angular momentum state for which $n_m = \langle
c^{\dag}_m c_m\rangle$ is nonzero. A compact droplet would have its last
occupied state occur at $m_{ed}\approx \nu^{-1}N$. A short examination of
Table I shows that $m_{eff}>m_{ed}$ and the difference
$W=m_{eff}-m_{ed}$ is precisely the number of CFs inhabiting the higher
LLs. For example, 
$m_{eff}-m_{ed}=N/2$
for half-filled case while $N-N_0$ is also $N/2$. At half filling, we
have demonstrated that the actual occupation numbers in the tail region
$m_{ed}<m<m_{eff}$ is quite small\cite{yang}. A few cases we have examined
for
other filling fractions also show similar behavior. This is consistent
with our plot of zeroes since we expect the electron
probabilities to be small in regions with high density of zeroes. 

The tail region $W$ is non-uniform in density, and there is no reason to 
expect 
incompressibility in this region. It is appropriate to regard the tail 
as a 2D metallic state which, like the ordinary 2D Fermi surface, supports a 
macroscopic number of particle-hole excitations. The abundance of low-energy 
modes in our wavefunctions can thus be understood as primarily excitations in 
the tail. 

We have found evidence of a Fermi liquid behavior in the occupation
profile of a half-filled droplet. 
\begin{figure}[t]
\centering
\epsfig{file=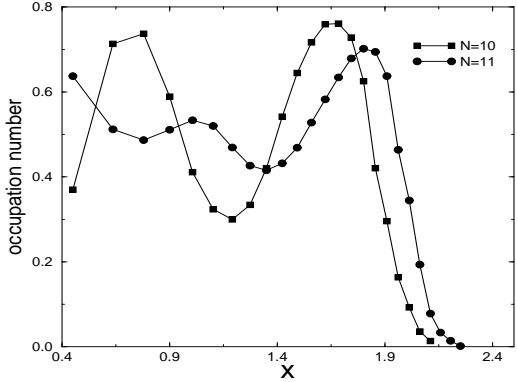,height=8cm,width=6cm,angle=270}
  \caption{The occupation profile of $N=10, 11$ droplet for half filling.
    The oscillatory period is close to one, indicating a $2k_F$ behavior.}
\label{fig:2k_F}
\end{figure}
\noindent
In Fig.\ \ref{fig:2k_F}, we plot the
occupation
numbers of $N=10$ and $11$ droplet as a function of
$x=\sqrt{2(m+1)}/\pi$. Both plots contain a nearly full 
oscillation with a period close to one. The choice of our variable $x$
follows from the analogy with the free Fermi gas, where the density
fluctuation in the presence of impurities is known to contain a term
$\cos (2k_F r+\delta)$. In our case, $k_F=\sqrt{2\nu}/l_{0}$ equals one
when the magnetic length $l_0$ is set to one, and we choose the average
radius of the single particle state 
$\sqrt{\langle  m| r^2 |m\rangle}l_0 =\sqrt{2(m+1)}$ to be $r$. The
periodicity close to one strongly hints that some characteristics of an
ordinary Fermi liquid is also apparent in the strongly interacting system
at $\nu=1/2$\cite{jkg}. 

In conclusion, we proposed here a class of ground state wavefunctions for
the droplet in the LLL, written entirely in terms of the
holormophic coordinates. A long and small-density occupation tail is a
generic
feature of the states, which is also reflected in the distribution of the
zeroes in the wavefunction. The zeroes of the wavefunction for general
filling factors are displaced
from the particles, except those governed by Fermi statistics. 
We have observed a $2k_F$-like oscillation in
the density profile at $\nu=1/2$ which we ascribe to the Fermi liquid
nature of this state. We have shown that the $\delta M_{tot}=1$ CM motion is 
exactly given as a linear superposition of the intra-Landau-level excitations 
of the CFs. 

J.H.H. acknowledges helpful discussions with David
Thouless, Wick Haxton and Jainendra Jain. 
S.R.E.Y. has been supported by
the Ministry of Education
under Grant No. BSRI-96-2444.

\begin{table}
\caption{Proposed wavefunctions for various filling factors 
$\nu=n/(2n\pm 1)$.}
\begin{tabular}{cccc}
$\nu$   & $N$    &  $\psi$   &      $m_{eff}$   \\ \hline
$2/5$   & $4k$   & [$3k$,2,1,..,1] &   $11k-3$ \\
        & $4k+1$ & [$3k+1$,1,..,1] &     $11k$  \\
$3/7$   & $6k$   & [$4k$,2,1,..,1] &   $16k-3$  \\
        & $6k+1$ & [$4k+1$,1,..,1] &     $16k$   \\ 
        & $6k+3$ & [$4k+2$,2,1,..,1]   & $16k+5$ \\
        & $6k+4$ & [$4k+3$,1,..,1]     & $16k+8$ \\
$1/2$   & $2k$   & [$k$,2,1,..,1]      & $5k-3$  \\
        & $2k+1$ & [$k+1$,1,..,1]    &    $5k$   \\
$3/5$   & $6k$   & [$2k$,2,1,..,1] &      $14k-3$  \\
        & $6k+1$ & [$2k+1$,1,..,1] &   $14k$     \\
        & $6k+3$ & [$2k+1$,2,1,..,1] &   $14k+4$ \\
        & $6k+4$ & [$2k+2$,1,..,1]   &   $14k+7$ \\
$2/3$   & $4k$   & [$k$,2,1,..,1] &   $9k-3$ \\
        & $4k+1$ & [$k+1$,1,..,1] &    $9k$
\end{tabular}
\end{table}

\widetext

\begin{references} 
\bibitem{books} For reviews, see {\it The Quantum Hall Effect}, edited 
by R. E. Prange and S. M. Girvin (Springer-Verlag, 1987), and {\it
Perspectives in Quantum Hall
Effects}, edited by S. Das Sarma and A. Pinczuk (John Wiley \& Sons, New
York, 1997).
\bibitem{drop} We will drop the universal factor 
$\prod_i e^{-|z_i |^2 /4}$ in this paper. 
\bibitem{laughlin} R. B. Laughlin, Phys. Rev. Lett. {\bf 50}, 1395 (1983).
\bibitem{jain}J. K. Jain, Phys. Rev. Lett. {\bf 63},  199 (1989).
\bibitem{rr} E. Rezayi and N. Read, Phys. Rev. Lett. {\bf 72}, 900 (1994). 

\bibitem{gv} S. M. Girvin and T. Jach, Phys. Rev. B {\bf 29}, 5617 (1984).
\bibitem{read} N. Read, Semicond. Sci. Technol. {\bf 9}, 1859 (1994).
\bibitem{recent} R. Shankar and Ganpathy Murthy, Phys. Rev. Lett. {\bf
79}, 4437 (1997); D. H. Lee, Phys. Rev. Lett. {\bf 80} 4745 (1998); V.
Pasquier and F. D. M.
Haldane, cond-mat/9712169; B. I. Halperin and Ady Stern, cond-mat/9802061;
N. Read, cond-mat/9804294.
\bibitem{acta} B. I. Halperin, Helv. Phys. Acta {\bf 56},
75 (1983).
\bibitem{JK} J. K. Jain and R. K. Kamilla, Int. J. Mod. Phys. B {\bf 11},
2621 (1997). 
\bibitem{HG} A set of explicitly lowest-Landau-level wavefunctions on
the sphere was proposed earlier by Ginocchio and Haxton (Phys. Rev. Lett.
{\bf 77}, 1568 (1996)) which turns out to be equivalent to Jain and
Kamilla
construction. 
\bibitem{girvin} S. M. Girvin, Phys. Rev. B {\bf 29}, 6012 (1984).   
\bibitem{trugman} S. A. Trugman and S. Kivelson, Phys. Rev. B {\bf 31},
5280 (1985).
\bibitem{comment} The states in the LLL with the reversed magnetic field are
given by anti-holomorphic wavefunctions $\bar{z}^m$.

\bibitem{yang} S.-R. Eric Yang and J. H. Han, Phys. Rev. B {\bf 57},
                 R12681 (1998). 
\bibitem{cll} X. G. Wen, Int. J. Mod. Phys. B {\bf 6}, 1711 (1992); C. L.
Kane and M. P. A. Fisher, in {\it Perspectives in Quantum Hall Effects}
(Ref. 1).
\bibitem{macdonald}A. H. MacDonald, S.-R. Eric Yang, and M. D. Johnson,
Aust. J. Phys. {\bf 46}, 345 (1993).

\bibitem{jkg}R. K. Kamilla, J. K. Jain, and S. M. Girvin,
Phys. Rev. B {\bf 56}, 12411 (1997).

\end{references}
\end{document}